\documentclass[prl,twocolumn,superscriptaddress,showpacs,amsmath,amssymb]{revtex4}
\usepackage{graphicx}
\usepackage{dcolumn}
\usepackage[dvipsnames]{color}
\usepackage{bm}
\usepackage{amsmath,amssymb,amsfonts}
\usepackage[dvipsnames]{xcolor}
\usepackage{txfonts}

\begin{document}
\title{Searching for near-horizon quantum structures in the binary black-hole stochastic gravitational-wave background}
\author{Song Ming Du}
\email{smdu@caltech.edu}
\affiliation{Theoretical Astrophysics and Walter Burke Institute for Theoretical Physics, California Institute of Technology, Pasadena, California 91125, USA}
 
\author{Yanbei Chen}
\affiliation{Theoretical Astrophysics and Walter Burke Institute for Theoretical Physics, California Institute of Technology, Pasadena, California 91125, USA}
\date{\today}

\begin{abstract}
Quantum gravity corrections have been speculated to lead to modifications to space-time geometry near black hole horizons.  Such structures may reflect gravitational waves, causing {\it echoes} that follow the  main gravitational waves from binary black hole coalescence. By studying two phenomenological models of the near-horizon structures under Schwarzschild approximation, we show that such echoes, if exist, will give rise to a stochastic gravitational-wave background, which is very substantial if the near-horizon structure has a near unity reflectivity for gravitational waves, readily detectable by Advanced LIGO.  In case the reflectivity is much less than unity, the background will mainly be arising from the first echo, with a level proportional to the power reflectivity of the near-horizon structure, but robust against uncertainties in  the location and the shape of the structure --- as long as it is localized and close to the horizon.  Sensitivity of  third-generation detectors allows the detection of a  background that corresponds to power reflectivity $\sim 3 \times 10^{-3}$, if uncertainties in the binary black-hole merger rate can be removed. We note that the echoes do alter the $f^{2/3}$ power law of the background spectra at low frequencies, which is rather robust against uncertainties.
\end{abstract}

\maketitle

\noindent {\it Introduction.--}  Black holes (BH) are monumental predictions of general relativity (GR)~\cite{Frolov}.  It is often believed that, inside a BH, a singularity exists, around which classical GR will break down and must be replaced by a full quantum theory of gravity (QTG).   The Planck scale of  $l_\text{P} \sim 1.6\times 10^{-35}\ \text m$ is often cited as the scale at which full-blown QTG is required.  However, interesting effects already arise as one applies quantum mechanics to fluctuations around the BH horizon, the boundary of the region from which one can escape toward infinity, even though space-time curvature does not blow up here.  Hawking showed that BHs evaporate, leading  to the so-called Black-Hole Information Paradox.   During attempts to resolve this Paradox --- as well as in other contexts --- it was proposed that space-time geometry near the horizon may differ from the Kerr geometry, by having additional, quantum structures~\cite{Giddings}.  Candidate proposals include firewall~\cite{Firewall}, fuzzball~\cite{Fuzzball} and gravastar~\cite{Gravastar}.

Detection of gravitational waves (GW) generated by binary black-hole (BBH) collisions marked the dawn of GW astronomy \cite{Abbott}, and brings an experimental tool to study the nature of BH horizon. Cardoso et al.\ proposed that geometric structures very close to the horizon can be probed by GW echoes that follow BBH waves, arising from the reflection from these structures, and the subsequent rebounds between these structures and the BH potential barrier~\cite{Cardoso,cardoso2017tests}. Whether the observed individual GW events have already provided positive experimental evidence towards the echoes  is still under debate ~\cite{Abedi, Abedi2,Ashton}. Furthermore, the particular echo model employed by ~\cite{Abedi, Abedi2} was considered rather naive and needed refinement~\cite{Price, Maselli}.  For example, Mark {\it et al.}, using scalar field generated by a point particle falling into a Schwarzschild BH, illustrated that, the echoes can have a variety of time-domain features, which depend on the location, and (in general frequency-dependent) reflectivity of the near-horizon structure \cite{Zach}. Echo structure during the entire inspiral-merger-ringdown wave was also analyzed in the Dyson series formalism in Ref.~\cite{correia2018characterization}.

In this letter, we propose to search for near-horizon structures via the stochastic GW background (SGWB) from BBH mergers.  Because the echo contribution to the background depends only on their energy spectra, it is much less sensitive to details of echo generation,  making the method more robust against uncertainties in the near-horizon structures.  We estimate the magnitude and rough feature of this SGWB, and illustrate its dependence on the near-horizon structure, following an Effective One-Body (EOB) approach: the two-body dynamics and waveform is approximated by the plunge of a point particle toward a Schwarzschild BH, following a trajectory that smoothly transitions from inspiral to plunge~\cite{ Buonanno,  Buonanno2}.

\noindent {\it GW amplitudes and power emitted.--} GWs emitted from a test particle plunging into a Schwarzschild BH can be described by the Sasaki-Nakamura (SN) equation \cite{Sasaki}:
\begin{align}
\left( \partial_{r_*}^2 +  \omega^2 - V_l(r) \right)X_{lm}(\omega, r_*)=S_{lm}(\omega,r), \label{SN}
\end{align}
where $r_*$ is the tortoise coordinate with $dr/dr_*=1-2M/r$  with effective potential given by 
\begin{align}
&V_l(r) = \left(1-\frac{2M}{r}\right)\left( \frac{l(l+1)}{r^2}-\frac{6M}{r^3}\right). \label{Vl}
\end{align}
with $M$ the mass of the BH. The source term is given by $S_{lm}(\omega, r) = W_{lm}(\omega, r) r^{-5} e^{-i\omega r_*}$, where $W_{lm}$ is a functional of the trajectory of the test particle and its explicit expression can be found in Eqs.~(19)---(21) of \cite{Sasaki}. The wave function $X_{lm}$ is related to GW in the $r\rightarrow +\infty$ limit via $h_+ + i h_\times = 8 r^{-1} \sum_{lm}  \ _{-2}Y_{lm} X_{lm}(t)$, where $_s Y_{lm}$ are spin-$s$ weighted spherical harmonics and $X_{lm}(t) = \int_{-\infty}^{+\infty}d\omega\ e^{-i\omega t} X_{lm}(\omega)$. The GW energy spectrum is given by
\begin{align}
{dE}/{d\omega} = \sum_{lm} 16\pi\omega^2 |X_{lm}(\omega, r_* \rightarrow \infty)|^2 \label{Energy}. 
\end{align}
For BHs, imposing  in-going boundary condition near the horizon and out-going condition near null infinity, solution to  Eq.~\eqref{SN} is expressed as  $X_{lm}^\text{(0)}(\omega, r_* \rightarrow \infty)  =e^{i\omega r_*} Z^{(0)}_{lm}(\omega)$, with
\begin{align}
Z_{lm}^{(0)}(\omega) = \int^{+\infty}_{-\infty}dr'_* \left[{S_{lm} (\omega,r_*')X_\text{in}^{(0)}(\omega,r_*')}\right]/{W^{(0)}(\omega)}, 
\label{Green}
\end{align}
with $W^\text{(0)}= X_\text{in}^{(0)}\partial_{r_*}X_\text{out}^{(0)} - X_\text{out}^{(0)}\partial_{r_*}X_\text{in}^{(0)}$ the Wronskian between the two homogenous solutions, with $X_\text{in}^{(0)} \sim e^{-i\omega r_*}$ for $r_* \rightarrow -\infty$ and $X_\text{out}^{(0)} \sim e^{+i\omega r_*}$ for $r_* \rightarrow +\infty$, respectively. 

\noindent {\it Echoes from near-horizon structure.--}  Let us now modify the Schwarzschild geometry near the horizon by creating a {\it Planck-scale potential barrier} $V_p$:  $ V_l \rightarrow V_l +V_p$, with $V_p$ centered at $r^p = 2M+\epsilon$,  with $\epsilon\ll M$; In tortoise coordinate, $\epsilon = l_p$ corresponds to $ r_*^p \approx - 182 M$. As discussed by \cite{Zach} the effect of $V_p$ is the same as replacing the  horizon ($r_*\rightarrow -\infty$) boundary condition for Eq.~(\ref{SN}) by 
\begin{align}
X^\text{(R)}_\text{in} \sim e^{-i\omega r_*} + \mathcal{R} e^{i\omega r_*} \qquad \text{for} \quad r_* \rightarrow r_*^p, \label{Boundary}
 \end{align}
 while keeping the  $r_*\rightarrow +\infty$ boundary condition unchanged.  Here,  $\mathcal{R}(\omega)$ can be viewed as a complex reflectivity of the potential barrier~\footnote{Here we will obtain $\mathcal{R}(\omega)$ from $V_p$, while the problem of obtaining $V_p$ once $\mathcal{R}(\omega)$ is measured is the so-called inverse scattering problem, see e.g., K.\ Chadan, {\it Inverse problems in quantum scattering theory}, Springer Science \& Business Media, 2012.}, the location of reflection is implicitly contained in its frequency dependence; for example, a Dirichlet boundary condition corresponds to $\mathcal{R}_D(\omega) =-e^{-2i\omega r_*^p}$~\footnote{In general, if $\mathcal{R}(\omega)= \rho(\omega) e^{i\psi(\omega)}$ with $\rho(\omega)$ a slowly varying complex amplitude and $\psi$ a fast-varying phase, then  the effective location of reflection for a wavepacket with central frequency $\omega_0$ is around $[\partial\psi/\partial\omega]_{\omega=\omega_0}/2$. }
.

\begin{figure}[t]
\centering
\includegraphics[width=50mm]{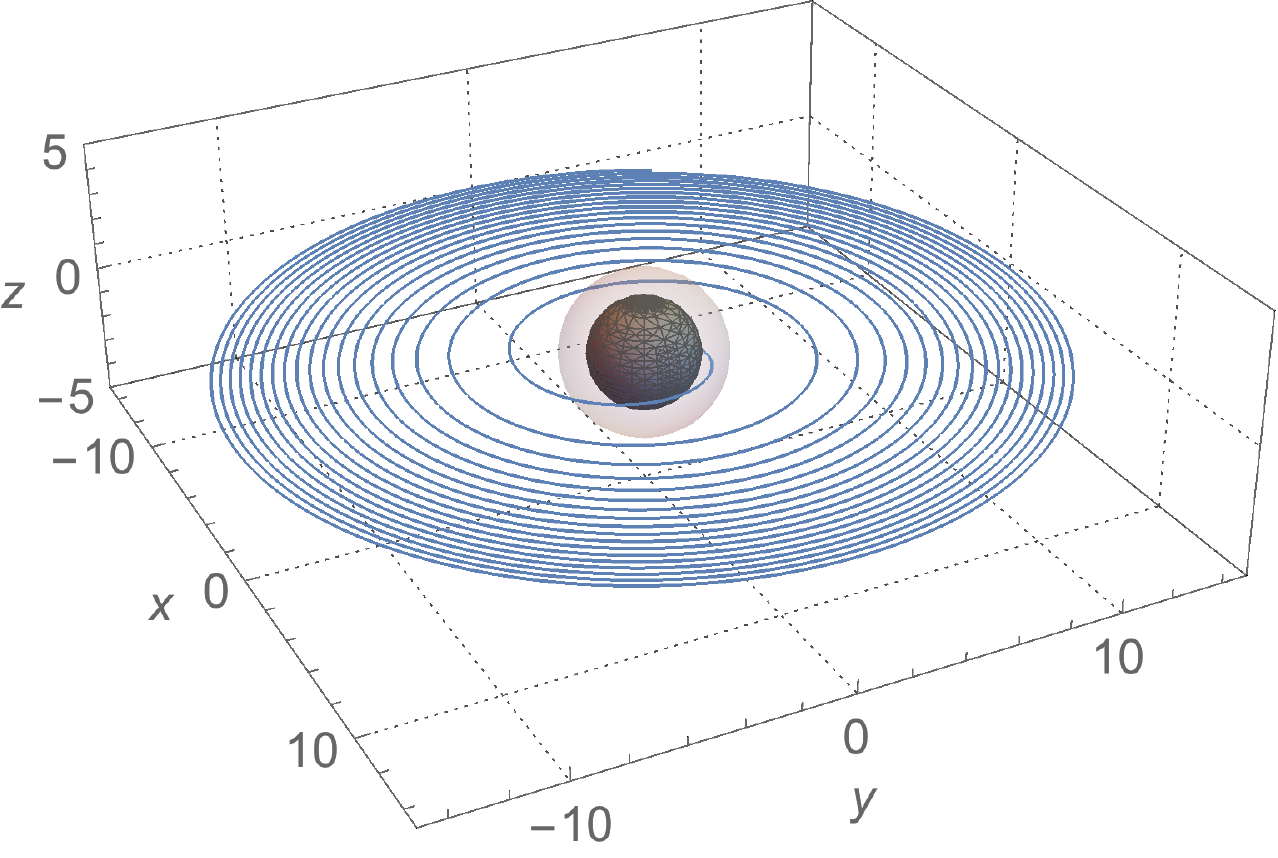}
\caption{Trajectory of the EOB effective particle moving in a coalescing quasi-circular orbit. The symmetric mass ratio $\nu = 0.25$. The inner black sphere with radius $2M$ represents the horizon of a Schwarzschild BH. The outer translucent sphere with radius $3M$ represents the photon sphere. }
\label{fig2}
\end{figure}
 
Defining $X_{lm}^\text{(R)} =Z_{lm}^\text{(R)} e^{i\omega r_*}$, $Z_{lm}^\text{(R)}$  can be written as a sum the main wave (for BH) and a series of echoes \cite{Zach}: 
\begin{align}
Z_{lm}^\text{(R)} =Z_{lm}^{(0)} +  \mathcal{R} {Z^{(1)}_{lm}}\sum_{n=0}^{+\infty} (\mathcal{R}\mathcal{R}_{\rm BH})^n,
\label{eqecho}
\end{align} 
with $\mathcal{R}_{\rm BH}$ the complex reflectivity of the Regge-Wheeler potential $V_l$ [see Eq.~(2.14) of \cite{Zach}] and 
\begin{align}
 Z^{(1)}_{lm} (\omega) = \int^{+\infty}_{-\infty}dr'_* \frac{S_{lm} (\omega,r_*')
\bar X_\text{in}^{(0)}(\omega,r_*')}{W^{(0)}(\omega)} +\mathcal{R}_{\rm BH} Z_{lm}^{(0)}, 
\label{Greenbar}
\end{align}
with $\bar X_\text{in}^{(0)}$ the complex conjugate of  $X_\text{in}^{(0)}$.

Note that each echo delayed from the previous one by  $\sim 2|r_*^p|$ in the time domain.  For small $\mathcal{R}$, we write 
$
Z_{lm}^\text{(R)} \approx Z^{(0)}_{lm} + \mathcal{R} Z^{(1)}_{lm}
$
and
\begin{align}
\label{beating}
\left(\frac{dE}{d\omega}\right)_R \! \approx \!16\pi\omega^2 \sum_{lm}
\left[
\left|  Z^{(0)}_{lm}\right|^2 +
\left|\mathcal{R}  Z^{(1)}_{lm}\right|^2 +
2\mathrm{Re} (\mathcal{R}  Z^{(1)}_{lm} \bar Z_{lm}^{(0)})\right].
\end{align}
This is the sum of energies from {\it main wave}, the {\it first echo}, and the {\it beat} between the main wave and the first echo.  While the beat is linear in $\mathcal{R}$, it is highly oscillatory in $\omega$, since the main wave and the echo are well separated in the time domain.  

\begin{figure}[b]
\centering
\includegraphics[width=0.4\textwidth]{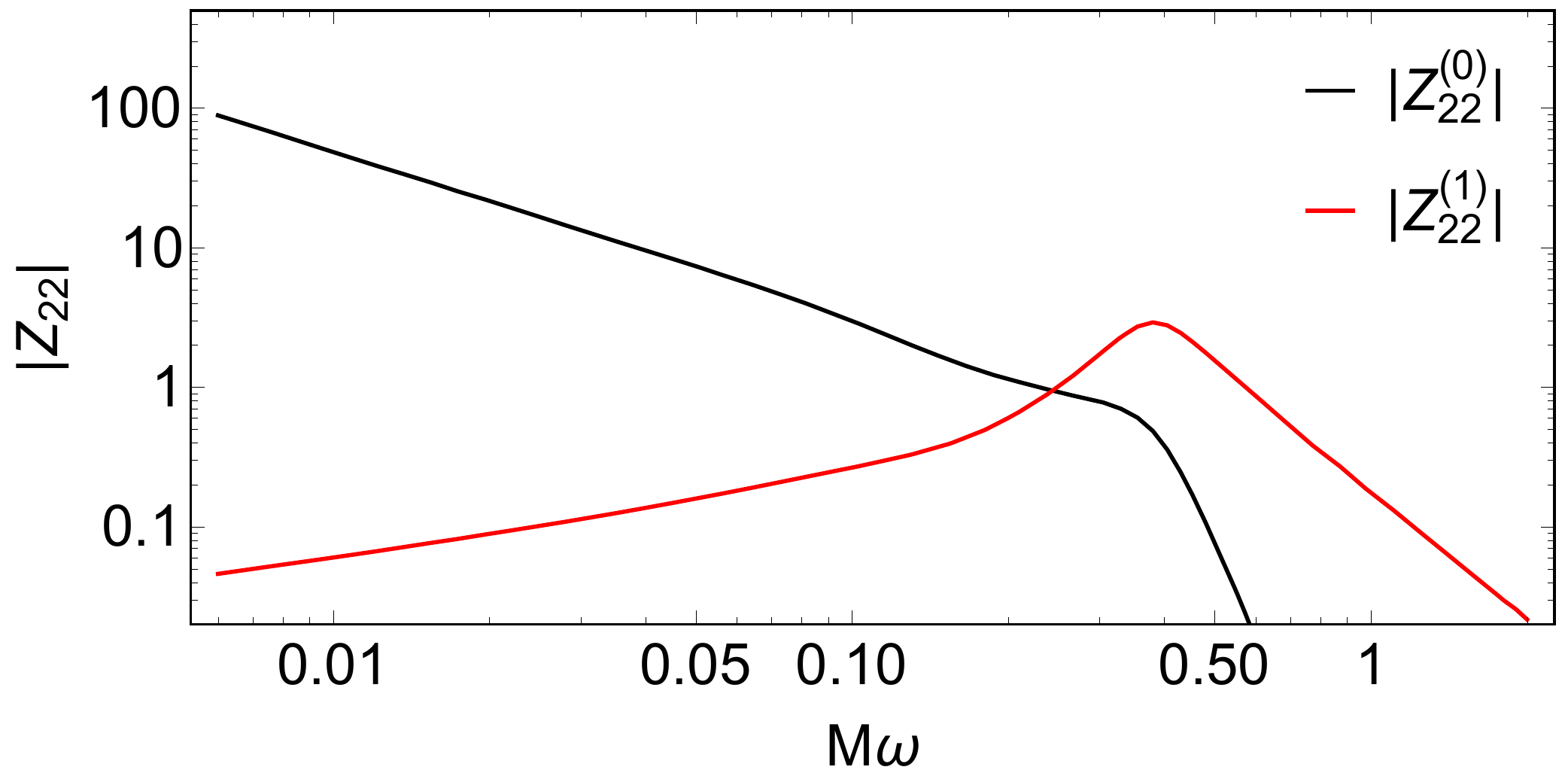}
\caption{The main wave $Z_{22}^{(0)}$ and the wave $Z_{22}^{(1)}$ that generates echoes via Eq.~\eqref{eqecho}. }
\label{figZ}
\end{figure}

\begin{figure*}[t]
\centering
\includegraphics[width=160mm]{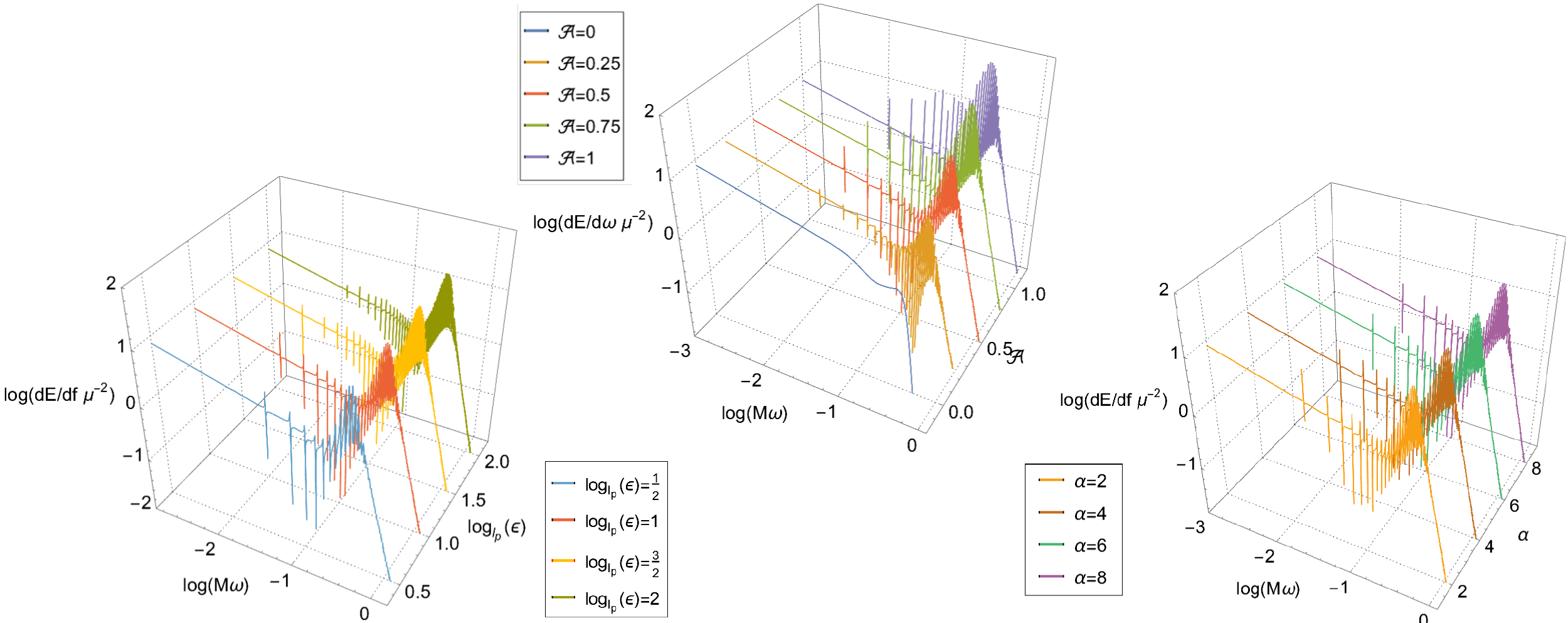}
\caption{
The energy spectra of GW emission from $\nu=0.25$ coalescing BBH. Upper Panel:  energy spectra for different values of $\mathcal{A}$, for $\epsilon=l_p$, with $\mathcal{R}$ given by Eq.~\eqref{Delta}. Left Panel: energy spectra for different values of $\epsilon$, for $\mathcal{A} = 0.5$ with $\mathcal{R}$ given by Eq. (\ref{Delta}) . Right Panel: energy spectra for different values of $\alpha$, for $\epsilon =l_p$, with $\mathcal{R}$ given by Eq. (\ref{Poschl}), fixing $\mathcal{A} = 2\alpha\lambda(1-\lambda) = 0.5$.}
\label{fig3}
\end{figure*}

 \noindent{\it Models of Reflectivity and Energy Spectra of Echoes.--} 
 Without prior knowledge about details of near-horizon structures, we only assume it is short-ranged and localized at $r_*^p$. The simplest would be to introduce a $\delta$-potential $V_p = \mathcal{A}\ \delta[(r_* - r_*^p)/M]$, with parameter $\mathcal{A}$ defined as the {\it area under the Planck potential}:  $\mathcal{A} = M\int_{-\infty}^{+\infty}  V_p\ dr_*$. Note that $\mathcal{A}$ is a dimensionless quantity.  As a comparison, the area under the Regge-Wheeler potential is \cite{Chandrasekhar} $M \int_{-\infty}^{+\infty}  V_l\ dr_* = (l-1)(l+2)/2 + 1/4$. Such a model corresponds to a reflectivity 
\begin{align}
\mathcal{R} (\omega) = e^{-2i\omega r_*^p} {\mathcal{A}}/({2 i M\omega -\mathcal{A})}\label{Delta}\,.
\end{align}
This is more physical than the Dirichlet case, by reducing $|\mathcal{R}|$ at larger $\omega$. Since $|\mathcal{R}(0) |=1$ and $\mathcal{R}(+\infty)= 0$ are general properties of  all physical potentials,  we expect Eq.~\eqref{Delta} to describe a large class of near-horizon quantum structures.
To further explore the shape of $V_p$, we also study the P{\"o}schl-Teller potential \cite{Poschl} $V_p=\alpha^2\lambda(1-\lambda)/M^2\cosh^{-2}[\alpha (r_*-r_*^p)/M]$. Dimensionless parameters $\alpha$ and $\lambda$ are related to the area under $V_p$ via $\mathcal{A}=2\alpha\lambda(1-\lambda)$. The corresponding reflectivity is \cite{Cevik}
\begin{align}
\mathcal{R}(\omega)=e^{-2i\omega r_*^p}\frac{\Gamma(i\frac{M\omega}{\alpha})\Gamma(\lambda-i\frac{M\omega}{\alpha})\Gamma(1-\lambda-i\frac{M\omega}{\alpha})}{\Gamma(-i\frac{M\omega}{\alpha})\Gamma(1-\lambda)\Gamma(\lambda)} \label{Poschl} \ ,
\end{align}
where $\Gamma(\cdot)$ is the Gamma function. In the following, we will keep $\mathcal{A}$ fixed and vary $\alpha$ and $\lambda$ to explore shapes of $V_p$.

To estimate of the echoes' energy spectrum, we adopt the EOB approach \cite{Buonanno,  Buonanno2}: for BHs with $m_1$ and $m_2$, 
we consider a point particle with reduced mass $\mu= m_1 m_2/(m_1+m_2)^2$ falling down a Schwarzschild BH with total mass $M=m_1+m_2$; the symmetric mass ratio is defined as $\nu = \mu/M$.   
For motion in the equatorial plane, we have a Hamiltonian for $(r,p_r,\phi,p_\phi)$, with radiation reaction incorporated as a generalized force $\mathcal{F}_\phi$ [Eqs. (3.41)--(3.44) of \cite{Buonanno2}]. 
Upon obtaining the trajectory (see Fig.~\ref{fig2} for $\nu=0.25$), we obtain source term $S_{lm}$, and compute $Z_{lm}^{(0)}$ and $ Z_{lm}^{(1)}$ using Eqs.~\eqref{Green} and \eqref{Greenbar}, which will then lead to the GW energy spectrum.

\begin{figure*}[t]
\begin{tabular}{ccc}
\includegraphics[width=0.45\textwidth]{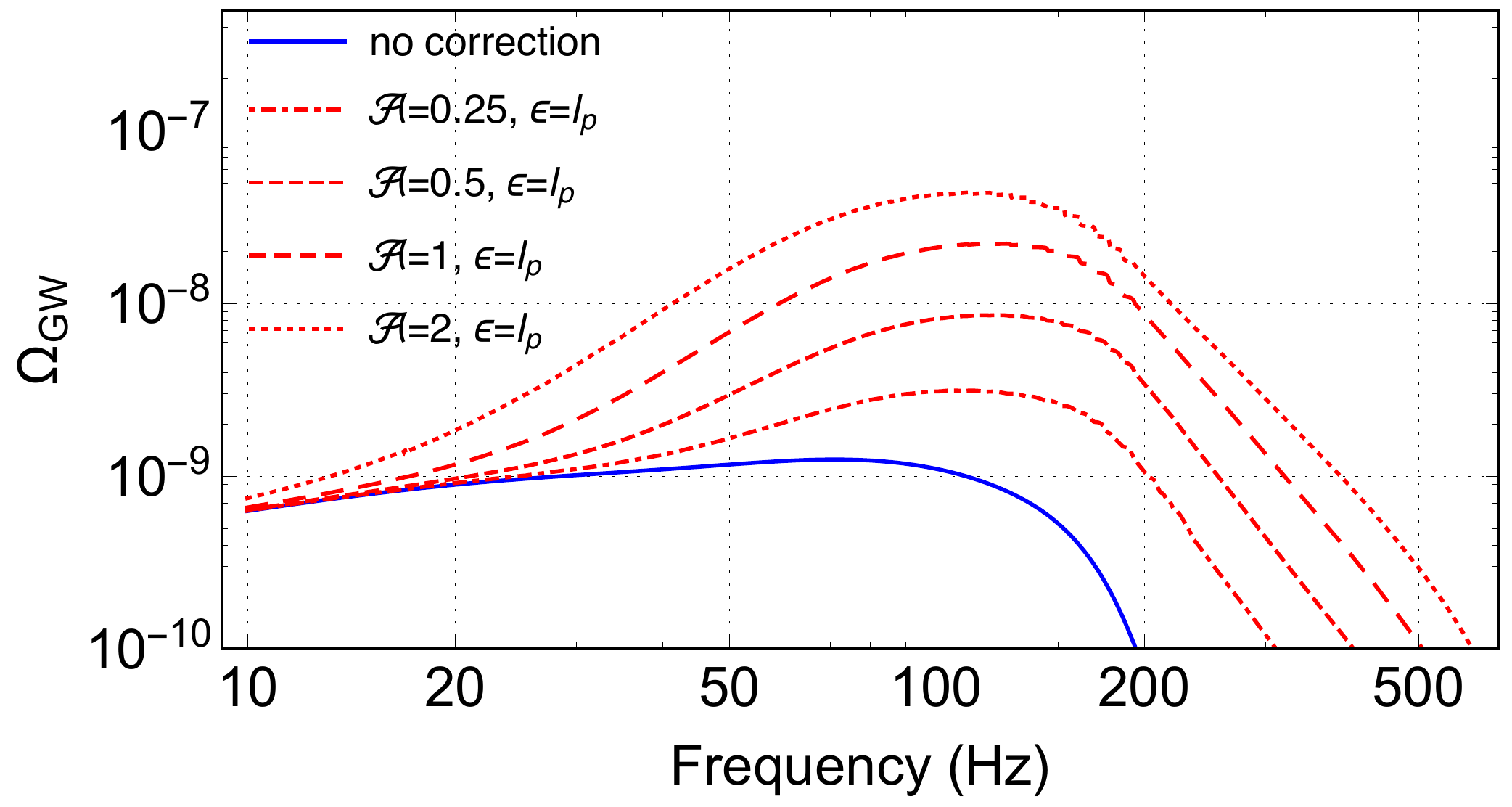} &
\hspace{0.75cm} &
\includegraphics[width=0.45\textwidth]{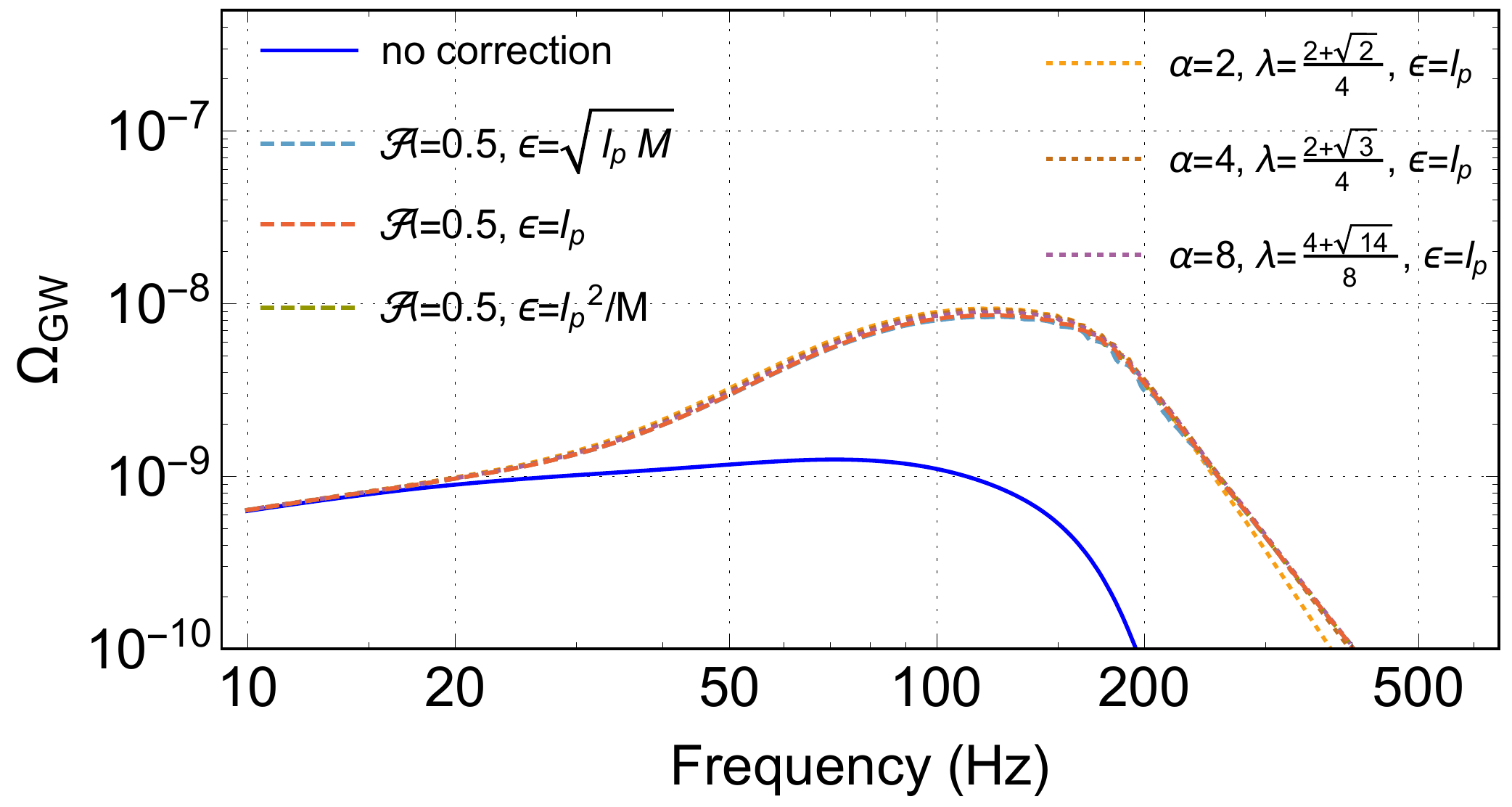}
\end{tabular}
\caption{
The influence to the fiducial \cite{Abbott2} BBH SGWB with varied areas, locations and shapes of the near-horizon potential. Left Panel: The spectral energy density $\Omega_\text{GW}(f)$ for different values of $\mathcal{A}$; reflectivity given by  Eq. (\ref{Delta}) with $\epsilon =l_p$. Right Panel: $\Omega_\text{GW}(f)$ for different values of $\epsilon$ (dashed) and $\alpha$ (dotted); reflectivity given by Eq. (\ref{Delta}) with $\mathcal{A}=0.5$ (dashed) as well as  Eq. (\ref{Poschl}) with $2\alpha\lambda(1-\lambda) = 0.5$, $\epsilon =l_p$ (dotted) respectively.
} 
\label{fig4}
\end{figure*}

We will focus on the $(l,m)=(2,2)$ mode, which carries most of the GW energy. 

As seen in Fig.~\ref{figZ},  the main wave $|Z_{22}^{(0)}|$ recovers the $f^{-7/6}$ power law at low frequencies,  as predicted by post-Newtonian approximation, also qualitatively mimics a BBH waveform at intermediate (merger) to high frequencies (ringdown).  Note that the ringdown makes the the $|Z_{22}^{(0)}|$  curve turn up slightly near the leading $(2,2)$ Quasi-Normal Mode (QNM) frequency of the Schwarzschild BH before sharply decreasing, similar to Fig.~3 of Ref.~\cite{Ajith}.  The wave $|Z_{22}^{(1)}|$ peaks roughly at the QNM frequency.

Horizon structures with $\mathcal{A}$ of order unity lead to significant modifications in GW energy spectrum $dE/d\omega$.  In the upper panel of Fig.~\ref{fig3},  we choose the reflectivity~\eqref{Delta} with $\epsilon=l_p$ and $\mathcal{A}=0.25$, 0.5, 0.75 and 1.  At low frequencies, near-horizon structures add peaks separated by 
$\Delta \omega \sim 0.017 M^{-1} \sim \pi/r_*^p$ to the post-Newtonian $dE/df\propto  f^{-1/3}$. These resonant peaks are related to the poles of $1/(1-\mathcal{R}\mathcal{R}_\text{BH})$ in the series sum of Eq.~\eqref{eqecho}. Near the QNM frequency, there is substantial additional radiation, which is due to the large value of $|Z_{22}^{(1)}|$. In the left panel, we choose several different values of $\epsilon$ which lead to different peak separation at low frequencies. In the right panel, we consider  reflectivity~\eqref{Poschl} and find that the shape of the Planck potential, as characterized by $\alpha$, has negligible influence to $dE/d\omega$ as long as the area keeps fixed.

\noindent {\it Stochastic Gravitational-Wave Background (SGWB).--} The SGWB is usually expressed as $\Omega(f) = {\rho_c}^{-1} {d\rho_\text{GW}}/{d\ln f} $, where $\rho_c$ represents the critical density to close the universe and $\rho_{GW}$ the GW energy density; it  is related  the $dE/df$ of a single GW source via \cite{Zhu}
\begin{align}
\Omega(f) = \frac{f}{\rho_c} \int_0^{z_\text{max}}dz\ \frac{R_m(z) [{dE}/{df}]_{f_z}}{(1+z)H(z)}, \label{Omega}
\end{align} 
where $f_z = f(1+z)$ is the frequency at emission. Here we adopt the $\Lambda$CDM cosmological model with $H(z) = H_0[\Omega_M(1+z)^3+\Omega_\Lambda]^{1/2}$, where the Hubble constant $H_0=70\text{km/s\ Mpc}$, $\Omega_M = 0.3$ and $\Omega_\Lambda = 0.7$. $R_m(z)$ is the BBH merger rate per comoving volume at redshift $z$. We use the {\it fiducial model} described in \cite{Abbott2}, where $R_m(z)$ is proportional to the star formation rate with metallicity $Z<Z_\odot/2$ and delayed by the time between BBH formation and merger. As in the Fiducial model, the parameters of BBH follow GW150914: $M=65M_\odot$, $\nu=0.25$ with a local merger rate $R_m(0) = 16 \text{Gpc}^{-3}\ \text{yr}^{-1}$.

\begin{figure}[t]
\centering
\includegraphics[width=0.4\textwidth]{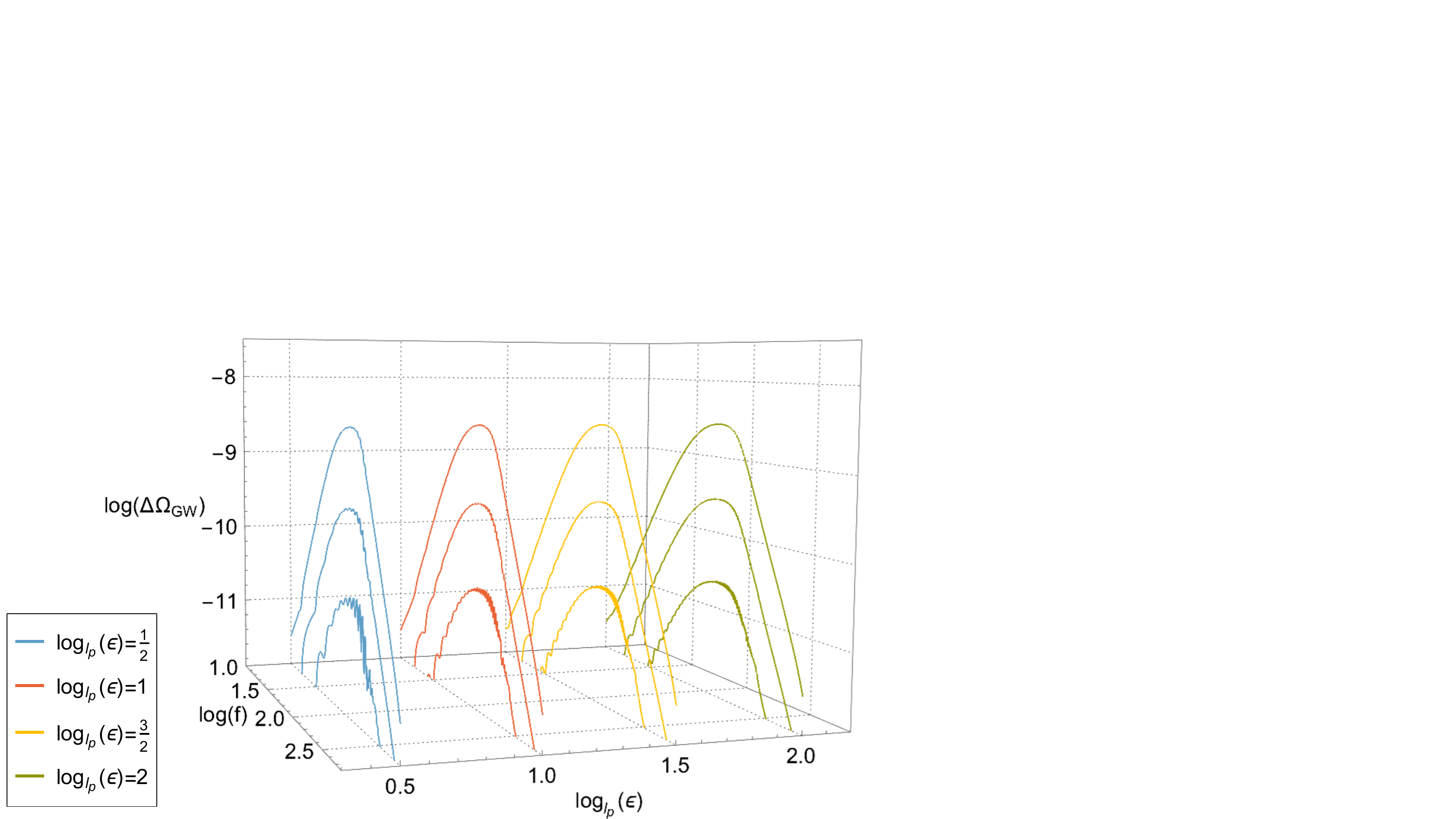}
\caption{$\Delta\Omega_{\rm GW}$ as functions of $f$, for $\mathcal{A}=0.3$, $0.1$ and $0.03$, and $\epsilon/M= (l_\mathrm{P}/M)^{1/2,1,3/2,2}$, reflectivity~(\ref{Delta}). Here $\Delta\Omega \propto \mathcal{A}^2$, except for the oscillations shown for small values of $\mathcal{A}$ and larger values of $\epsilon$, when the beat term of Eq.~\eqref{beating} is not completely smoothed out. Other parameters are the same as in the fiducial model \cite{Abbott2}.}
\label{figX}
\end{figure}

For $\mathcal{A} \sim 1 $, we get substantial additional SGWB from the echoes (left panel of  Fig. \ref{fig4}) in a way that is insensitive to the location and shape of the near-horizon structure, as characterized by $\epsilon$ and $\alpha$ (right panel). This robustness  indicates the area under the Planck potential is the most relevant observable of the near-horizon structures in SGWB.  For smaller $\mathcal{A}$, we plot the {\it additional} SGWB,  defined as $\Delta\Omega \equiv \Omega_{\mathcal{A}>0} -\Omega_{\mathcal{A}=0}$ in Fig.~\ref{figX}. Here $\Delta\Omega$ is approximately  $\propto\mathcal{A}^2$, for $\mathcal{A} > 0.03$ and $\epsilon /M < \sqrt{l_{\rm P}M}$: beating between the main wave and the echoes Eq.~\eqref{beating} is unimportant, and the additional SGWB mainly arise from energy contained in the first echo.  

\noindent {\it Detectability.--}
The optimal signal-to-noise ratio (SNR) for a SGWB between a pair of detectors is given by $ \sqrt{\langle \Omega|\Omega\rangle }$ \cite{Callister}, with 
\begin{align}
\langle \Omega_\text{A}|\Omega_\text{B}\rangle \equiv 2T\left(\frac{3H_0^2}{10\pi^2}\right)^2\int_0^{+\infty}df\ \frac{\Omega_\text{A}(f)\gamma^2(f)\Omega_\text{B}(f)}{f^6P_1(f)P_2(f)}, \label{Product}
\end{align}
where $\gamma(f)$ is the normalized overlap reduction function between the detectors, and  $P_{1,2} (f)$ are the detectors' noise spectral densities.  We consider  advanced LIGO at design  sensitivity \cite{Ajith}, LIGO Voyager \cite{Voyager} and Einstein Telescope (ET) \cite{Sathyaprakash} at planned sensitivities. Advanced LIGO and LIGO Voyager have the same $\gamma$ and we take the constant  $\gamma = -3/8$ for co-located ET detectors~\cite{Nishizawa}. The 1-year SNRs are listed in Table \ref{SNR} for values of $\mathcal{A}$ at order unity, in which case the echoes contribute significantly to the SNRs.
 \begin{table}[ht] 
\begin{tabular}{ccccc} 
\hline \hline 
$\mathcal{A}$ & LIGO  & Voyager & ET \\ 
\hline
0 &  1.42 &  27.5  &  196 \\ 
0.25 & 1.60 & 30.8 & 270 \\ 
0.5 & 2.15 & 40.9 & 513 \\ 
1 & 3.99 & 75.2 & 1215 \\ 
2 & 8.76 & 164.7 & 2561 \\ 
\hline \hline 
\end{tabular}
\caption{One-year SNR of three generations of GW detectors for SGWB $\Omega_\mathcal{A}$, varying $\mathcal{A}$. The reflectivity corresponds to Eq. (\ref{Delta}) with $\epsilon=l_p$. Other parameters are the same as in the fiducial model \cite{Abbott2}.} \label{SNR} 
\end{table}

For lower values of $\mathcal{A}$,  we apply the model-selection method of Ref.~\cite{Callister} to distinguish the SGWB with and without echo contributions. The log-likelihood ratio (LR) between two models is given by $\ln{\Lambda}=\langle \Delta \Omega|\Delta \Omega\rangle/4$ and two models considered discernible when  $\ln{\Lambda} > c >1$.  Here we choose $c=12$, which corresponds to a false alarm rate of $10^{-6}$ \cite{Wilks}.  Minimum distinguishable  $\mathcal{A}$ to reach this LR threshold is shown in Tab.~\ref{Distinguishable}; with 5-year integration, Voyager can detect $\mathcal{A} \approx 0.21$, while ET can detect $\mathcal{A}\approx 0.042$.

\begin{table}[ht] 
\begin{tabular}{ccccc} 
\hline \hline 
$T$ & LIGO  & Voyager & ET \\ 
\hline
1 yr &  $1.87$ &  $0.32$  &  $0.062$ \\ 
5 yrs & $1.07$ & $0.21$  &  $0.042$ \\ 
\hline \hline 
\end{tabular}
\caption{ The minimal distinguishable $\mathcal{A}$ to reach a log-likelihood ratio $\ln{\Lambda} > 12$ for current and future GW detector with different integration times. The reflectivity corresponds to Eq. (\ref{Delta}) with $\epsilon=l_p$. Other parameters are the same as in the fiducial model \cite{Abbott2}.} \label{Distinguishable} 
\end{table}

\noindent {\it Conclusions and Discussions.--}  As we have seen in this paper, the $\Delta\Omega$ due to the echoes is largely independent from uncertainties in $r_*^p$.  For strong near-horizon structures, with $\mathcal{A}$ the order of unity, SGWB  from the echoes will be clearly visible. For weak near-horizon structures, $\Delta\Omega$ is mainly given by the first echo, and is simply proportional to the power reflectivity $|\mathcal{R}|^2$. The level detectable by ET corresponds to  $\mathcal{A} \sim 0.042$, which corresponds to $|\mathcal{R}|^2\approx3\times10^{-3}$ near the peak of the echo energy spectrum. Further details of the background not only depends on details in the Planck potential barrier $V_p$, we will also need to generalize the analysis to a Kerr BH. 

Uncertainties also exist in the SGWB of the main, insprial-merger-ringdown wave, e.g., arising from different star formation rates, different metallicity thresholds to form BHs, details in the evolution of binary stars and the distributions in the time delay between BBH formation and merger --- all of these lead to uncertainties in the local BBH merger rate and the local distribution of mass $M$ and symmetric mass ratio $\nu$~\cite{Abbott2}. It is believed these uncertainties will be well quantified and  narrowed down by future BBH detections. For example, the range of BBH local merger rate has been narrowed down to $12 - 213 \text{ Gpc}^{-3} \text{ yr}^{-1}$ using GW170104 \cite{Abbott3}. On the other hand, as demonstrated by Zhu {\it et al.}, these uncertainties only scale the background spectra linearly at low frequencies and hence keep the power law $\Omega(f) \propto  f^{2/3}$ for $f < 100~\text{Hz}$ unchanged \cite{Zhu}. Our result shows the appearance of the near-horizon structures changes the slfaope  of $\Omega(f)$, making it devaite from the $f^{2/3}$ power law even at low frequencies.  This may be used to alleviate the influence from uncertainties. 

In addition to BBH, binary neutron star (BNS) mergers also contribute to the background with a comparable magnitude \cite{Abbott4}. Within the bandwidth of ground-based GW detectors, this background arises solely from inspiral, which gives an $f^{2/3}$ power law and is not influenced by the presence of the near-horizon structure. As a result, the echo SGWB $\Delta \Omega$ remains unchanged and our analysis on detectability still holds.

Echoes may also be detectable from individual events. Our calculations indicate for an event similar to GW150914, to reach an echo SNR of 10 the value of $\mathcal{A}$ should be at least $0.24$ (LIGO), $0.050$ (Voyager) and $0.011$ (ET), respectively. However, in the matched filtering search of individual signal, the exact waveform is required, which in our model depends not only on $\mathcal{A}$, but also on $\epsilon$ and $\alpha$, 
but may depend further on other unknown details of the Planck-scale potential --- making it less robust. An analysis combined both background and individual signals will be presented in a separate publication\cite{Du}.

\begin{acknowledgements}
\noindent {\it Acknowledgements.--}  This work is supported by NSF Grants PHY-1708212 and PHY-1404569 and PHY-1708213. and the Brinson Foundation. We thank Yiqiu Ma, Zachary Mark, Aaron Zimmerman for discussions, in particular ZM and AZ for sharing insights on the echoes. We are grateful to Eric Thrane and Xing-Jiang Zhu for providing feedback on the manuscript.
\end{acknowledgements}

\bibliography{references}

\end{document}